\documentclass[conference]{IEEEtran}

\usepackage[noadjust]{cite}

\usepackage{array,colortbl,multirow,multicol,hhline,booktabs,ctable}

\ifCLASSINFOpdf
\else
\fi
\usepackage{amssymb}
\usepackage[cmex10]{amsmath}
\usepackage{algorithmic}

\usepackage{array}

\usepackage{mdwmath}
\usepackage{mdwtab}

\usepackage[tight,footnotesize]{subfigure}
\usepackage{url}

\begin{document}
%
\title{A Potential Approach to Overcome Data Limitation \\in Scientific Publication Recommendation}


\author{
\IEEEauthorblockN{Hung Nghiep Tran\IEEEauthorrefmark{1}, Tin Huynh\IEEEauthorrefmark{2}, Kiem Hoang\IEEEauthorrefmark{3}}
\IEEEauthorblockA{University of Information Technology, VNU-HCMC, Vietnam\\Email: \IEEEauthorrefmark{1}nghiepth@uit.edu.vn, \IEEEauthorrefmark{2}tinhn@uit.edu.vn, \IEEEauthorrefmark{3}kiemhv@uit.edu.vn}
}

\maketitle

\begin{abstract}
Data are essential for the experiments of relevant scientific publication recommendation methods but it is difficult to build ground truth data. A naturally promising solution is using publications that are referenced by researchers to build their ground truth data. Unfortunately, this approach has not been explored in the literature, so its applicability is still a gap in our knowledge. In this research, we systematically study this approach by theoretical and empirical analyses. In general, the results show that this approach is reasonable and has many advantages. However, the empirical analysis shows both positive and negative results. We conclude that, in some situations, this is a useful alternative approach toward overcoming data limitation. Based on this approach, we build and publish a dataset in computer science domain to help advancing other researches.
\end{abstract}


%
\IEEEpeerreviewmaketitle

\section{Introduction}
As in other areas, data play a key role in relevant scientific publication recommendation research. The purpose of this task is to recommend publications that are relevant to researchers' research interests \cite{sugiyama2010scholarly}. In this task, the most popular evaluation approach is offline evaluation, which is based on ground truth data that contain the publications which are known to be relevant to each researcher. Building ground truth data is usually difficult and expensive.

Some researchers manually built ground truth data by surveying a group of researchers, which is costly so the datasets are usually small and prone to biased evaluation. Moreover, due to privacy issues, these datasets are not fully shared. Some other researchers use datasets adapted from reference management systems but their offline and online evaluation results have been shown to be conflicting \cite{beel2013comparative}. Therefore, it is crucial to find another approach to build ground truth data.

Naturally thinking, since researchers cite publications that are related to their researches, references are relevant to researchers' interests. Hence, it is intuitive to built ground truth data based on reference data. A similar approach has been used in the task of recommending citation, a.k.a. citation prediction. Unfortunately, to the best of our knowledge, this approach has not been explored in the task of recommending relevant publication. In this paper, we systematically study this approach to give some insights about its applicability.

First, we construct and analyze the hypotheses based on the information needs theory to support this approach. In addition, we propose a process to automatically build ground truth data from bibliographic data. Then, we empirically assess it by statistical analysis of the evaluation results on two types of ground truth data, the automatically-built and the manually-built. We show that this approach is reasonable and has many advantages. However, the empirical analysis shows both positive and negative results. In general, evaluation on two types of ground truth data are almost consistent but detailed analysis shows that it is not always confidently guaranteed.

This research presents the following main contributions.
\begin{enumerate}
  \item We systematically explore the approach that builds ground truth data for evaluation of relevant scientific publication recommendation based on reference data. We conclude that, in some situations, this is a reasonable and useful alternative approach toward overcoming data limitation.
  \item We propose a specific process to build ground truth data from bibliographic data with many advantages.
  \item We build and publish a dataset in computer science domain to help advancing research in scientific publication recommendation.
\end{enumerate}

Sect. 2 presents a summary of related research. In Sect. 3, we theoretically analyze the approach and Sect. 4 describes the process to build ground truth data. Sect. 5 presents experimental settings and Sect. 6 shows the results. Sect. 7 concludes.

\section{Related Work}
The emerging trend in relevant scientific publication recommendation research has raised an increasing attention to ground truth data \cite{sugiyama2010scholarly,le2014scirecsys,ohta2011related}. Many approaches have been proposed but all of them have flaws. Some researchers manually build ground truth data by surveying a group of researchers. With a high cost, these datasets are usually small, e.g., with only 28 researchers \cite{sugiyama2010scholarly}. Moreover, due to privacy issues, these datasets are not fully shared. Some researchers use datasets adapted from reference management systems, e.g., Mendeley and Docear. However, ground truth data in these datasets are not guaranteed to represent researchers' real research interests. Another approach to build dataset is crowdsourcing but it has not been used in relevant scientific publication recommendation area and requires complicated systems.

A promising approach is building ground truth data based on reference data. However, to the best of our knowledge, no research has been done on assessing its applicability. In this paper, we systematically explore this approach.

\section{Hypotheses}
In this section, we construct and analyze the hypotheses to support the approach to build ground truth data based on reference data.

\subsection{Information Needs}
The purpose of recommending relevant publications for researchers is to satisfy researchers' information needs by providing publications that match their research interests. Information needs is the desire to obtain information to satisfy a conscious or unconscious need \cite{taylor1962process}. In this research, we analyze researchers' information needs in the context of doing research and writing scientific publications w.r.t. research phases \cite{chu1999literary}. Most researches start with a literature survey, then developing new ideas and doing experiment, finally writing publications to report the results, and publishing them.

Ideally, in such a process, researchers seek potentially relevant publications and select which are worth reading. After reading, they can identify relevant publications. When researchers write their publications, they cite the most relevant publications. Citing is the activity that shows an official acknowledgment of researchers to a publication as a fulfillment for their information needs. Information needs could be expressed in different ways and citing is the one that actually expose the information needs of researchers. Based on this analysis, we construct the following hypothesis.

\textbf{Hypothesis 1.} \textit{In the context of doing research and writing scientific publications, there are many levels of exposing a researcher's information needs in which citing is the highest level.}

\subsection{Relevant Publications}
When conducting a research, researchers are gradually realizing their information needs, which are finally expressed by citing a set of references. These references are equivalent to information requirements in the research context \cite{taylor1962process}. Ideally, these references really do satisfy the researchers' information needs, either consciously or unconsciously. Thus, they are relevant to the researchers. We note that researchers could not cite all publications that are relevant, so these references are not exhaustive but they are the most important ones.

Naturally, researchers may temporally change their research interests, so relevancy is temporal, i.e., references form the set of relevant publications in a specific period of time. Future references, which are newly cited in the future but not in the past, reflect the changes in researchers' research interests. This type of information provides an interesting instrument for evaluation. Finally, we come up with the following hypothesis.

\textbf{Hypothesis 2.} \textit{References made by researchers are their relevant publications. Moreover, they are the most important ones.}

\subsection{Evaluation Ability}
Evaluation is essential in recommendation systems research \cite{shani2011evaluating}. Its function is to compare different methods. The most popular evaluation approach is offline evaluation \cite{shani2011evaluating} based on the preprepared relevant items in ground truth data. Thus, the most important property of ground truth data is relevancy.

With regard to reference data, researchers have spent a lot of time and effort to find the most relevant publications to cite. This is a serious process occurring in real life. Hence, the relevancy of references is actually judged based on their content by researchers, so, references have the potential to be used as ground truth data. However, ground truth data need to be unobserved at the training phase of experiments. To guarantee this condition, we could separate data into past data and future data based on publications' published year using a timestamp defined as the present. Training data are only extracted from past data, while ground truth data are from future data.

With regard to ordinary ground truth data, a popular approach to get relevant publications is surveying researchers. This process is similar to the process producing references. However, researchers usually lack time and motivation, so in some cases, relevancy might be decided based on publication titles only. As a result, it lacks the depth and coverage of reference data. So, reference data are theoretically more suitable to be used to build ground truth data. We summarize this analysis by the following hypothesis.

\textbf{Hypothesis 3.} \textit{Future references could be used as ground truth data in evaluation of recommending relevant publications for researchers.}

Online evaluation is another approach that is more expensive but considered better than offline evaluation \cite{shani2011evaluating}. This approach is based on the real interactions between users and systems to compare different methods. We believe that reference data, which are also based on real interactions, are potentially as good as online evaluation.

In this paper, we partially assess these hypotheses through the comparison with ordinary ground truth data. Extensive assessment, which requires comparison with online evaluation, is out of the scope of this paper and reserved for future research.

\section{Process to Build Dataset}

\subsection{Concepts}
These concepts are used in the proposed process to build ground truth data.

\textit{\textbf{Definition 1 (Timeline).}} Assuming we already have a dataset $D$. Considering the year of publication in $D$, timeline $T$ starts by the earliest published year $T_s$ and ends by the latest published year $T_e$ with a step by one year in a linear timescale.

\textit{\textbf{Definition 2 (Present, Past, and Future).}} We select a year $T_0 \in T$ as the \textit{Present}. Then, the \textit{Past} is $[T_s, T_0)$. The \textit{Future} is $[T_0, T_e]$. We also define the \textit{Restricted Past} $[T_p, T_0), T_p \in [T_s, T_0)$, which could be a small period. Similarly, we define the \textit{Restricted Future} $[T_0, T_f], T_f \in [T_0, T_e]$.

\textit{\textbf{Definition 3 (Target Researcher).}} \textit{Target Researcher Set} $R$ contains those ones for whom recommendations are generated. First, we define target researcher $r$ as those who actively publishes at least $n_p$ and $n_f$ publications in the \textit{Restricted Past} and the \textit{Restricted Future} w.r.t. a specific timeline $T$, respectively. Then, $R$ is the set of target researcher $r$.

\textit{\textbf{Definition 4 (Candidate Publication).}} Candidate publication set $P$ is a set of publications from which recommendations are selected. They must contain items in ground truth data. Thus, these publications must be published before $T_0$. Other restricting conditions may be applied to reduce the size of $P$. These conditions are specified later.

\textit{\textbf{Definition 5 (Reference Data).}} With the defined timeline, for each researcher $r \in R$, considering publication $p$ that is published before $T_0$. If $r$ has not cited $p$ in the \textit{Restricted Past} but cite $p$ in the \textit{Restricted Future}, we say $p$ is a \textit{Future Reference}. For each researcher $r$, the set of publications that satisfy the above condition is the \textit{Future Reference Set} $FR_r$.

\textit{\textbf{Definition 6 (Ground Truth Data).}} For each researcher $r \in R$, her \textit{Ground Truth Data} $GT_r$ contain her relevant publications w.r.t. a specific timeline $T$. In our approach, the \textit{Future Reference Set} $FR_r$ is used as \textit{Ground Truth Data} $GT_r$. To distinguish between the this type of ground truth data and the ordinary ground truth data, the latter is denoted as $OGT_r$.

\subsection{Bibliographic Data}
The dataset $D$ and its ground truth data are built based on bibliographic data. Hence first, we need a bibliographic dataset $BD$. Some researches have been done on building bibliographic data \cite{huynh2012integrating}. These data usually contain noises, especially author name ambiguity \cite{ferreira2012brief}. Many approaches to solve this problem have been proposed recently \cite{ferreira2012brief, tran2014author}.

\subsection{Building the Dataset}
\label{process}
Given bibliographic dataset $BD$, we define the timeline $T$ to separate data into the \textit{Restricted Past} and the \textit{Restricted Future} using \textit{\textbf{Def. 1 \& 2}}. Then, we identify the target researchers using \textit{\textbf{Def. 3}}. The \textit{Target Researcher Set} $R$ is selected uniformly randomly from target researchers. For each researcher $r \in R$, we extract \textit{Future Reference} data using \textit{\textbf{Def. 5}}. Then they are used as \textit{Ground Truth Data} $GT_r$ for researcher $r$ as in \textit{\textbf{Def. 6}}. To reduce the size of candidate publication set as in \textit{\textbf{Def. 4}}, we propose merging publications from ground truth data of all \textit{Target Researchers} and to use as candidate publications, i.e., $P = \cup_{r \in R} \{p, p \in GT_r\}$.

\section{Empirical Assessment Settings}
\subsection{Datasets}
\subsubsection{The Ordinary Dataset $D'$}
For comparison, we use the manually built ordinary dataset $D'$ with the \textit{Ordinary Ground Truth Data} built by surveying 15 junior researchers \cite{sugiyama2010scholarly}.

\subsubsection{The Proposed Dataset $D$}
Dataset $D$ are built as described in Sect. \ref{process}. The bibliographic dataset $BD$ is from Microsoft Academic Search. $BD$ contains publications in computer science domain from 1951 to 2010. First, we have the timeline $T$ with the \textit{Present} set at $2006$, the \textit{Restricted Past} $[2001, 2006)$, and the \textit{Restricted Future} $[2006, 2010]$. Similarly to $D'$, we also select junior researchers that have published the first time in the \textit{Restricted Past}, with 1 or 2 publications in the \textit{Restricted Past} and at least 5 publications in the \textit{Restricted Future}. Then, we sample uniformly randomly 100 researchers as \textit{Target Researcher Set} $R$. Finally, the statistics of $D$ are shown in Table \ref{tab:data} together with $D'$.

\begin{table}[htbp]
  \centering
  \caption{Datasets.}
    \begin{tabular}{r|r|r}
    \hline
    Dataset & $D$   & $D'$ \\
    \hline
    Target Researchers & 100   & 15 \\
    Written Publications per Researcher & 1.3   & 1.3 \\
    Citations per Publication & 2.7   & 0 \\
    References per Publication & 10.2  & 18.7 \\
    Candidate Publications & 4023  & 597 \\
    Relevant Publications per Researcher & 42.8  & 28.6 \\
    \hline
    \end{tabular}%
  \label{tab:data}%
\end{table}%

\subsection{Evaluation Results Consistency}
We empirically assess this approach by statistically analyze the correlation between evaluation results on $D$ and $D'$. We use evaluation results of \textit{Content-based Filtering (CBF)} methods because only \textit{CBF} results are available on dataset $D'$ \cite{sugiyama2010scholarly}. In most \textit{CBF} methods, publications and researchers are represented as feature vectors to compute their similarities, then a number of publications with highest similarities are selected to recommend.

For each publication $p$, we can compute \textit{tf-idf} vector $V_p$ from its content \cite{sugiyama2010scholarly}. Publication $p$ may have references $pr$ and citations $pc$, which are cited by $p$ and citing $p$, respectively. We can combine \textit{tf-idf} vector of $p$ with $pr$ and $pc$ to build other kinds of feature vector. In general, there are four kinds of feature vector $F_p$:

\begin{enumerate}
\item Based on its own content ($F1$): \\$F_p = V_p$.
\item Combining with its references ($F2$): \\$F_p = V_p + \sum V_{pr}$.
\item Combining with its citations ($F3$): \\$F_p = V_p + \sum V_{pc}$.
\item Combining with its references and citations ($F4$): \\$F_p = V_p + \sum V_{pr} + \sum V_{pc}$.
\end{enumerate}

Similarly, feature vector of researcher $r$ could be computed by combining feature vectors of her publications in the past $rp$. That is, $F_r = \sum F_{rp}$.

Different \textit{CBF} methods are constructed by combining different kinds of feature vector. Hence, there could be maximally 16 distinct \textit{CBF} methods. However, due to data limitation on $D'$, there are only two kinds of researcher's feature vector based on $F1$ and $F2$. As a result, there are eight distinct \textit{CBF} methods on $D'$. The similarity between researcher and publication is computed by the popular cosine similarity $cos(F_p,F_r)$ \cite{sugiyama2010scholarly}. The evaluation results of these methods are computed using three common metrics, \textit{NDCG@5}, \textit{NDCG@10}, and \textit{MRR} to get two sets of evaluation results denoted as $eval_D$ and $eval_{D'}$ on each dataset $D$ and $D'$, respectively \cite{sugiyama2010scholarly}.

The main experiments are the statistical analysis of evaluation results on two datasets. We compute the correlations between $eval_D$ and $eval_{D'}$ using three popular correlation coefficients: Pearson's $r$, Spearman's $\rho$, and Kendall's $\tau$. Since a method does not necessarily outperforms other methods on all metrics, it is necessary to compare the evaluation results on all metrics. Thus, first, we compute the correlations between $eval_D$ and $eval_{D'}$ on all methods and all evaluation metrics. Then we compute the correlations on each metric separately to assess the approach in more detail.

\section{Results and Discussions}
\subsection{Evaluation Results Consistency}
Table \ref{tab:cbfresult} shows the evaluation results of different \textit{CBF} methods given by combining different kinds of feature vector of publication and researcher on row and column, respectively. Results on D' are reported by Sugiyama and Kan \cite{sugiyama2010scholarly}.

\begin{table}[htbp]
  \centering
  \caption{Results of different CBF methods.}
    \begin{tabular}{c|c|r|r|r|r|r}
    \hline
    \multicolumn{3}{c|}{\multirow{3}[6]{*}{\textit{CBF} results}} & \multicolumn{4}{c}{Researcher feature vector} \\
    \hhline{~~~----}
    \multicolumn{3}{c|}{}  & \multicolumn{2}{c|}{Dataset $D$} & \multicolumn{2}{c}{Dataset $D'$} \\
    \hhline{~~~----}
    \multicolumn{3}{c|}{}  & F1    & F2    & F1    & F2 \\
    \hline
    \multirow{12}[6]{*}{Publication} & \multirow{4}[2]{*}{NDCG@5} & F1    & 0.347 & 0.365 & 0.382 & 0.442 \\
    \multirow{12}[6]{*}{feature vector} &       & F2    & 0.338 & 0.351 & 0.402 & 0.405 \\
          &       & F3    & 0.375 & 0.359 & 0.388 & 0.429 \\
          &       & F4    & 0.387 & 0.388 & 0.418 & 0.445 \\
    \hhline{~------}
          & \multirow{4}[2]{*}{NDCG@10} & F1    & 0.298 & 0.324 & 0.392 & 0.401 \\
          &       & F2    & 0.299 & 0.316 & 0.403 & 0.399 \\
          &       & F3    & 0.335 & 0.318 & 0.401 & 0.406 \\
          &       & F4    & 0.343 & 0.348 & 0.407 & 0.403 \\
    \hhline{~------}
          & \multirow{4}[2]{*}{MRR} & F1    & 0.502 & 0.461 & 0.455 & 0.505 \\
          &       & F2    & 0.457 & 0.479 & 0.453 & 0.494 \\
          &       & F3    & 0.503 & 0.472 & 0.450 & 0.477 \\
          &       & F4    & 0.503 & 0.498 & 0.472 & 0.538 \\
    \hline
    \end{tabular}%
  \label{tab:cbfresult}%
\end{table}%

Table \ref{tab:correlation} shows the correlation of evaluation results between two datasets. If \textit{$\mid$coefficient value$\mid$} $< 0.2$ or \textit{p-value} is large, the result is reported as no correlation in the table.

\begin{table}[htbp]
  \centering
  \caption{Correlation coefficients value.\newline$***, \dag, \ddag$ denote significant levels 0.001, 0.2, 0.3, respectively.}
    \begin{tabular}{l|c|c|c}
    \hline
    Coefficient & Pearson & Spearman & Kendall \\
    \hline
    General correlations & 0.48$^{***}$ & 0.81$^{***}$ & 0.64$^{***}$ \\
    \hline
    On NDCG@5 & 0.61$^{\ddag}$ & 0.60$^{\dag}$ & 0.50$^{\dag}$ \\
    On NDCG@10 & No    & 0.54$^{\dag}$ & 0.43$^{\dag}$ \\
    On MRR & No    & No    & No \\
    \hline
    \end{tabular}%
  \label{tab:correlation}%
\end{table}%

\subsection{Discussions}
Theoretical analysis shows that the approach to build ground truth data based on references is reasonable. It also suggests that this approach has the potential to give more useful evaluation results than ordinary ground truth data.

Datasets built based on this approach have some valuable properties. First, they contain rich data, so they are suitable for many methods, e.g., publication content is used in \textit{CBF}, \textit{Past Reference} is used as rating in \textit{CF}, etc. Second, it is practically easy to extend the dataset and select the desired properties of data, e.g., junior researchers. For experiments, we built dataset $D$ with the properties compatible with the ordinary dataset $D'$. They both contain junior researcher; Table \ref{tab:data} shows that they are almost similar. As shown in Table \ref{tab:cbfresult}, the evaluation results on $D$ are reasonable and meaningful.

Comparison of evaluation results on two datasets shows some interesting insights. In general, evaluation results on all metrics between two datasets are statistically significant strong positive correlated, with high coefficient and very low significant level. High correlations suggest that evaluation results on two datasets are almost consistent. Very low \textit{p-value} $<$ 0.001 guarantees that the correlations are statistically generalizable. Thus, we could use $D$ instead of $D'$ for evaluation in some situations, e.g., roughly selecting some recommendation methods before conducting an expensive online evaluation.

However, evaluation results on separate metrics are less correlated. Particularly, the correlation is still strong on \textit{NDCG} but there is almost no correlation on \textit{MRR}. This means the order of recommended items is consistently evaluated but the position of the first correct ones is not. The high \textit{p-value} suggests that increasing the number of samples in the comparison might give more confident results. On the other hand, theoretical analysis shows that this approach provides some advantages over ordinary ground truth data, so comparison with offline evaluation on ordinary dataset may not be enough for fully assessment. This suggests that comparisons with online evaluation is needed to further explore this approach.

\section{Conclusion}
Data are essential for the experiments of relevant scientific publication recommendation methods but it is difficult to build ground truth data. A naturally promising solution is building ground truth data based on reference data. Unfortunately, this approach has not been explored in the literature. In this research, we systematically study this approach.

First, we construct and analyze the hypotheses to support this approach. Then, we empirically assess this approach by a statistical analysis. We show that this approach is reasonable and has many advantages. However, the empirical analysis shows both positive and negative results. We conclude that this approach is useful in some situations. In addition, we propose a process to build ground truth data from bibliographic data. Based on this approach, we build and publish a dataset in computer science domain to help advancing other researches.

The theoretical analysis suggests that this approach is a potential solution toward overcoming data limitation. Future work should focus on extensive assessment of this approach, especially by comparison with online evaluation.

\section*{Acknowledgment}
We wish to thank Prof. Atsuhiro Takasu for the inspiring discussions. This research is funded by Vietnam National 
University HoChiMinh City (VNU-HCM) under grant number C2014-26-03.

\bibliographystyle{ieeetran}

\end{document}